\documentclass[11pt]{article}
\usepackage{epsfig,sint,macros,cite}
\begin{document}
\begin{titlepage}
\begin{flushright}
  CERN-TH-2000/175\\
  DESY-00-090
\end{flushright}

\vskip 0.5 cm
\begin{center}
  {\Large\bf 
  Effective Chiral Lagrangians and Lattice QCD  \\[0.5ex] }
\end{center}
\vskip 0.5 cm
\vbox{
\centerline{
\epsfxsize=2.5 true cm
\epsfbox{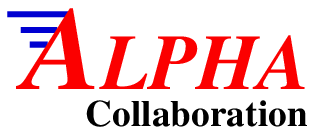}}
}
\vskip 0.5 cm
\begin{center}
{\large 
     Jochen Heitger$^{\scriptscriptstyle a}$,
     Rainer Sommer$^{\scriptscriptstyle a}$ and
     Hartmut Wittig$^{\scriptscriptstyle b,}$\footnote{On leave of
     absence from: Theoretical Physics, Oxford University, 1 Keble
     Road, Oxford OX1~3NP, UK}
\vskip 0.5cm
$^{\scriptstyle a}$
DESY \\
Platanenallee 6, D-15738 Zeuthen, Germany
\vskip 2.5ex
$^{\scriptstyle b}$
CERN, Theory Division,\\
CH-1211 Geneva 23, Switzerland
\vskip 3.0cm
{\bf Abstract}}
\vskip 0.7ex
\end{center}

We propose a general method to obtain accurate estimates for some of
the ``low-energy constants'' in the one-loop effective chiral
Lagrangian by means of simulating lattice QCD. In particular, the
method is sensitive to those constants whose values are required to
test the hypothesis of a massless up-quark. Initial tests performed in
the quenched approximation confirm that good statistical precision can
be achieved. As a byproduct we obtain an accurate estimate for the
ratio of pseudoscalar decay constants, $\Fk/\Fpi$, in the quenched
approximation, which lies 10\% below the experimental result. The
quantities that serve to extract the low-energy constants also allow a
test of the scaling behaviour of different discretizations of QCD and
a search for the effects of dynamical quarks.

\vfill

\begin{center}
June 2000
\end{center}

\eject

\vfill

\eject

\end{titlepage}

\section{Introduction \label{sec_intro}}

Chiral Perturbation Theory (ChPT) \cite{chir:GaLe1,chir:GaLe2} plays
an important r\^ole in the study of low-energy phenomena in QCD. As is
well known, ChPT is based on simultaneous expansions of scattering
amplitudes or hadronic matrix elements in powers of momentum and the
quark masses $\mup,\,\mdown$ and $\mstrange$, and the form of the
effective chiral Lagrangian is entirely determined by chiral symmetry.
Another feature is the emergence of a number of coupling constants at
every given order, which incorporate the non-perturbative dynamics of
QCD. These couplings -- sometimes called ``low-energy constants'' --
are {\em a priori\/} undetermined and can be fixed through
phenomenological constraints in conjunction with additional
assumptions. Currently the typical accuracy which is achieved in the
determination of the low-energy constants is not very high (for a
review, see~\cite{chir:param}).

It has been noted some time ago that the mass parameters
$\mup,\,\mdown$ and $\mstrange$ cannot be fixed unambiguously from
symmetry considerations alone. The reason is that the effective chiral
Lagrangian is invariant under a family of transformations of the quark
masses, which can be absorbed into the low-energy
constants~\cite{chir:KapMan86}. In particular, this hidden symmetry of
the chiral Lagrangian implies that one cannot distinguish between
$\mup=0$ and $\mup\approx5\,\MeV$, whilst preserving all other
phenomenological predictions. Thus, additional theoretical input
beyond ChPT is required to decide whether $\mup=0$ is indeed realized
in nature. An obvious strategy is then to sharpen the current
estimates for the low-energy constants and check whether they are
compatible with the condition that $\mup=0$.

Since $\mup=0$ represents a simple solution to the strong CP problem,
this question has continued to attract a lot of attention, despite
several plausible arguments, each of which independently rules out
$\mup=0$~\cite{chir:Leut90,chir:Leut96}. However, the problem has
never been studied in a first principles approach starting from the
QCD Lagrangian directly.

In this paper we propose and test a method to determine a large set of
low-energy constants with good accuracy using lattice simulations.
Given its technical feasibility, such an approach eliminates the use
of theoretical assumptions in the specification of the chiral
Lagrangian.  In the context of testing $\mup=0$ the r\^ole of lattice
calculations has recently been discussed by Cohen, Kaplan and
Nelson~\cite{chir:CoKapNel99}. We expand on their proposal by
incorporating the information gained by studying the mass dependence
of matrix elements in addition to that of the pseudoscalar masses.
Furthermore, we present ready-to-use numerical procedures which
increase the statistical precision and discuss the influence of
lattice artefacts. The latter point is of great relevance because the
effective chiral Lagrangian of Gasser and Leutwyler is not valid for
non-zero lattice spacing.

Our method is generally applicable in simulations of QCD with or
without dynamical quarks. This initial study mainly serves to test its
accuracy and limitations, and for that purpose it is sufficient to
apply it to lattice data obtained in the quenched approximation. As a
consequence we also consider comparisons of our lattice data with the
quenched version of ChPT, thereby extracting some of the low-energy
constants relevant for the quenched theory.

On the basis of our lattice data we conclude that the mass dependence
of matrix elements can be determined with high precision in lattice
simulations. Furthermore we show that low-energy constants in the
chiral Lagrangian can be obtained with a typical, absolute statistical
accuracy of $\pm\,0.05$ in the continuum limit. Systematic
uncertainties due to neglected higher orders are estimated to be
$\pm\,0.2$. However, a systematic study of the influence of higher
orders in the chiral expansions has so far not been performed, owing
to limitations in the range of light quark masses that one is
currently able to explore. Future calculations at smaller quark masses
will be required in order to settle this point.

The remainder of this paper is organized as follows. In
Section~\ref{sec_chpt} we recall the main concepts of ChPT. Our
computational strategy is described in Section~\ref{sec_stg}, and the
numerical details are discussed in Section~\ref{sec_contlim}, focusing
on the extrapolations to the continuum. Our results are presented in
Section~\ref{sec_results}, and Section~\ref{sec_concl} contains some
concluding remarks. Two appendices list some details about expressions
in partially quenched ChPT and explain our choice of additional
low-energy constants in the quenched theory.

\hyphenation{mass-re-nor-ma-li-za-tion}

\section{Chiral perturbation theory \label{sec_chpt}}

In this section we repeat the main features of ChPT and specify our
notation. In addition to standard ChPT we also discuss its quenched
and partially quenched versions.

\subsection{Standard ChPT and $\mup=0$}

The effective chiral Lagrangian is written as a low-energy
expansion~\cite{chir:GaLe1,chir:GaLe2}
\bes
  {\cal L}_{\rm eff} =
  {\cal L}_{\rm eff}^{(2)}+{\cal L}_{\rm eff}^{(4)}+\ldots 
\ees
The leading contribution ${\cal L}_{\rm eff}^{(2)}$ contains two
coupling constants $F_0$ and $B_0$. To lowest order $F_0$ coincides
with the pion decay constant $\fpi$. Throughout this paper we adopt a
convention in which $\fpi\approx93\,\MeV$. At order $p^4$ additional
couplings $\alpha_1,\alpha_2,\ldots,\alpha_{12}$ appear in the
effective chiral Lagrangian\,\footnote{We adopt a convention in which
  the $\alpha_i$'s are related to the corresponding couplings~$L_i$ of
  ref.~\cite{chir:GaLe2} through $\alpha_i=8(4\pi)^2{L_i}$.}. These
couplings are not constrained by chiral symmetry.  Furthermore they
contain the divergences that arise if one goes beyond tree level and
thus depend on the renormalization scale (and scheme). As mentioned in
the introduction, experimental information at low energies is not
sufficient to specify the values of the complete set of couplings
$\alpha_1,\ldots,\alpha_{12}$. For this reason one has to add further
theoretical constraints, which usually involve certain assumptions,
such as large-$\Nc$ arguments. By applying such a procedure, the
``standard'' values of the low-energy constants in our convention for
$\Nf=3$ flavours
read~\cite{chir:GaLe1,chir:GaLe2,chir:BijCor88,chir:Bij90,chir:RiGaDoHo91,chir:Babu92,chir:param}
\bes
\begin{array}{lcl}
   \alpha_1 =\sm0.2 \pm0.4 & \qquad & \alpha_6 = -0.5\pm0.4  \\
   \alpha_2 =\sm1.07\pm0.4 & \qquad & \alpha_7 = -0.5\pm0.25  \\
   \alpha_3 =-4.4 \pm1.4   & \qquad & \alpha_8 = \sm0.76\pm0.4  \\
   \alpha_4 =-0.76 \pm0.6  & \qquad & \alpha_9 = \sm7.8\pm0.8  \\
   \alpha_5 =\sm0.5 \pm0.6 & \qquad & \alpha_{10} = -6.1\pm0.8.
\end{array} \label{eq_Li_stand}
\ees
Here the $\alpha_i$'s have been renormalized at scale
$\mu=4\pi{F_\pi}$, which will always be used in the remainder of this
paper. The constants $\alpha_{11}$ and $\alpha_{12}$ are of little
phenomenological interest and are not included here.

The question whether $\mup=0$ has been discussed at length in the
literature~\cite{chir:KapMan86,chir:Leut90,chir:Leut96,chir:CoKapNel99}.
The usual starting point is the observation that simultaneous
transformations of the quark masses
\bes
  \mup^\lambda=\mup+\lambda\mdown\mstrange,\quad
  \mdown^\lambda=\mdown+\lambda\mstrange\mup,\quad
  \mstrange^\lambda=\mstrange+\lambda\mup\mdown,
\ees
and coupling constants according to
\bes
  \alpha_6^\lambda = \alpha_6+\lambda\frac{(4\pi F_0)^2}{4B_0},\quad
  \alpha_7^\lambda = \alpha_7+\lambda\frac{(4\pi F_0)^2}{4B_0},\quad
  \alpha_8^\lambda = \alpha_8-\lambda\frac{(4\pi F_0)^2}{2B_0},
\label{eq_KMamb}
\ees
leaves the effective chiral Lagrangian invariant. Here, $\lambda$
denotes an arbitrary parameter, and for $\lambda\not=0,\,\mup=0$ one
can generate an effective up-quark mass such that all predictions by
ChPT, in particular for ratios of quark masses, remain unchanged. In
order to test whether $\mup=0$ one has to determine the linear
combination~\cite{chir:Leut90} $(2\alpha_8-\alpha_5)$, which, however,
is not directly accessible from phenomenology. The value of $\alpha_5$
is obtained from the ratio of kaon to pion decay constants,
$\Fk/\Fpi$, but $\alpha_8$ can only be inferred from the linear
combination
\bes
  \alpha_5-12\alpha_7-6\alpha_8,
\label{eq_L578}
\ees
which is related to the Gell-Mann--Okubo formula. Clearly this linear
combination is invariant under the transformation of~\eq{eq_KMamb}. As
pointed out in~\cite{chir:CoKapNel99} a choice for $\alpha_7$ and
$\alpha_8$ which is compatible with $\mup=0$ is given by
\bes
  \mup=0:\quad \alpha_8=-0.9\pm0.4,\quad
  \alpha_7=0.25\pm0.25,
  \label{eq_Li_m0}
\ees
which is quite different from the corresponding numbers listed in
\eq{eq_Li_stand}. The task for lattice simulations is now to provide
independent determinations of $\alpha_5,\,\alpha_8$ from linear
combinations which are not invariant under \eq{eq_KMamb}, starting
from the QCD Lagrangian. Provided that these estimates turn out to be
sufficiently accurate, it should then be possible to test the
hypothesis that $\mup=0$.

\subsection{Partially quenched ChPT}

The r\^ole of lattice simulations for the determination of the
$\alpha_i$'s has already been stressed
in~\cite{chir:CoKapNel99,ShaSho_lat99}, and most recently
in~\cite{ShaSho_L7}. In particular, it has been noted that simulations
of ``partially quenched'' QCD, in which the sea and valence quarks can
be chosen independently, may be useful as well as technically
feasible. Thus, one is not forced to simulate at the physical values
of the $\up$, $\down$ and~$\strange$ quarks. Instead, the only
requirement is that the pseudoscalar masses be small compared with the
typical chiral scale of $\Lambda_\chi\approx1\,\GeV$. Hence the use of
moderately light sea and valence quark masses and their independent
variation may be sufficient to extract the low-energy constants.

It has to be kept in mind, though, that the values of the $\alpha_i$'s
have to be determined separately for $\Nf=2$ and~3 flavours of
dynamical quarks. This is a relevant point since there is currently
not much experience with simulation algorithms for odd numbers of
flavour. 

Partially quenched ChPT has been studied to O$(p^4)$ by a number of
authors~\cite{chir:Sha97,chir:GolLeu97,ShaSho_lat99}. Here we focus on
the one-loop expressions for pseudoscalar masses and decay constants
obtained by Sharpe~\cite{chir:Sha97} for $\Nf$ {\em degenerate\/}
flavours of sea quarks. For the remainder of this paper we also take
over some of the notation used in that reference. In particular, we
denote the mass of the sea quark by $m_S$, whereas the masses of the
(in general non-degenerate) valence quarks are denoted by $m_1,\,m_2$.
As in~\cite{chir:Sha97} we introduce the dimensionless parameters
\bes
  y_{ij}=\frac{B_0(m_i+m_j)}{(4\pi F_0)^2},\qquad
  i,\,j \in \{1,\,2,\,S\}.
\label{eq_yparam_PQ}
\ees
By setting $\Lambda_\chi=4\pi F_0$ in eqs.~(14) and~(18) of
ref.~\cite{chir:Sha97}, we obtain the generic formulae for the
pseudoscalar mass and decay constant, i.e.
\bes
  \mp^2 &=& y_{12}(4\pi F_0)^2 
\bigg\{1+\frac{1}{\Nf}\left[\frac{y_{11}(y_{SS}-y_{11})
  \ln{y_{11}} -y_{22}(y_{SS}-y_{22})\ln{y_{22}}}{y_{22}-y_{11}}
  \right]  \nonumber\\
  & & +y_{12}(2\alpha_8-\alpha_5)
      +y_{SS}\Nf(2\alpha_6-\alpha_4)\bigg\} \label{eq_MP_pqchpt} \\
  \frac{\Fp}{F_0} &=& 1-\frac{\Nf}{4}\left( y_{1S}\ln{y_{1S}}
  +y_{2S}\ln{y_{2S}} \right) \nonumber \\
  & & +\frac{1}{2\Nf}\left(
  \frac{y_{11}y_{22}-y_{SS}y_{12}}{y_{22}-y_{11}}
  \ln\frac{y_{11}}{y_{22}}+y_{12}-y_{SS}\right) \nonumber \\
  & & +y_{12}\frac{1}{2}\alpha_5 +y_{SS}\frac{\Nf}{2}\alpha_4.
  \label{eq_FP_pqchpt}
\ees
Here the constants $\alpha_i$ are to be evaluated at scale
$\mu=\Lambda_\chi=4\pi F_0\approx1170\,\MeV$. These expressions will
later be used to form quantities that allow for the extraction of the
$\alpha_i$'s using lattice data.

\subsection{Quenched ChPT}

Quenched ChPT has originally been discussed in
refs.~\cite{chir:BerGol92,chir:Sha92}. The complete chiral Lagrangian
to order~$p^4$ in quenched QCD has been studied by Colangelo and
Pallante~\cite{chir:ColPal97}. Their results form the basis of our
analysis. 

As is well known flavour singlets have to be treated differently in
the quenched approximation: the decoupling of the $\eta^\prime$ from
the pseudoscalar octet by means of the anomaly is not realized in the
quenched theory. Therefore, the quenched chiral Lagrangian contains
additional low-energy constants associated with flavour-singlet
contributions. These include the singlet mass scale $m_0$ and the
coupling constant $\alpha_\Phi$, which multiplies the kinetic term of
the singlet field in the quenched chiral Lagrangian. The mass scale
$m_0$ is usually expressed in terms of the parameter~$\delta$ defined
by
\bes
  \delta = \frac{m_0^2}{3(4\pi F_0)^2}.
\ees
From various lattice studies (e.g.
\cite{eta:Kuramashi94,decay:BhatGup,qspect:CPPACS,eta:FNAL_lat99}) one
expects $\delta\approx0.1$. For $\alpha_\Phi$ the available estimates
have been summarized in~\cite{Sharpe_lat96} as
$\alpha_\Phi\approx0.6$.

Following ref.~\cite{chir:ColPal97} we restrict ourselves to the case
of degenerate quarks. The complete results at one loop for the
pseudoscalar mass and decay constant read\,\footnote{In
  ref.~\cite{chir:ColPal97} a different combination of low-energy
  constants appears in the expression for $\mp^2$, since the authors
  use an alternative convention for the
  counterterms~\cite{ColPal_pcomm}. The convention employed in this
  paper is consistent with that used in standard and partially
  quenched ChPT.}
\bes
  \mp^2 &=& y\,(4\pi F_0)^2\left\{
  1-\left(\delta-\textstyle\frac{2}{3}\alpha_\Phi{y}\right) 
   \left[1+\ln{y}\right]
   +\left((2\alpha_8^{\rm q}-\alpha_5^{\rm q})
  -\frac{\alpha_\Phi}{3}\right)y  \right\}   \label{eq_MP_qchpt}\\ 
   \frac{\Fp}{F_0} &=& 1+y\,\textstyle\frac{1}{2}\alpha_5^{\rm q}.
  \label{eq_FP_qchpt}
\ees
Here $y$ is defined as
\bes
   y=\frac{2B_0m}{(4\pi F_0)^2},
\ees
and $\alpha_8^{\rm q},\,\alpha_5^{\rm q}$ denote the analogues of the
low-energy constants $\alpha_8$ and $\alpha_5$ in the quenched theory.

\section{Strategy \label{sec_stg}}

We now introduce the procedure to determine the low-energy constants
from lattice data by studying the quark mass dependence of suitably
defined {\em ratios\/} of pseudoscalar masses and matrix elements. In
particular it is useful to investigate the mass dependence around some
reference quark mass $\mref$. It is important to realize that this
reference point does not have to coincide with a physical quark
mass~\cite{mbar:pap3}. We only require that it should lie inside the
range of simulated quark masses and within the domain of applicability
of ChPT.

\subsection{The ratios $\RFG$ and $\RF$}

Let us consider the case of degenerate valence quarks, $m_1=m_2=m$,
either in the quenched approximation or in partially quenched QCD at a
fixed value of the sea quark mass $m_S$. If we introduce
\bes
  y=\frac{2B_0m}{(4\pi F_0)^2},\quad 
  \yref=\frac{2B_0\mref}{(4\pi F_0)^2}, \quad x=y/\yref,
\ees
then the ratios defined by
\bes
  \RFG(x) &=& \left(\frac{\Fp(y)}{\Gp(y)}\right)\left/
  \left(\frac{\Fp(\yref)}{\Gp(\yref)}\right)
  \right. \label{eq_RFG_def1} \\[0.2cm]
  \RF(x)  &=& \Fp(y)\left/\Fp(\yref)\right.
\ees
are universal functions of the parameter~$x$, which measures the
deviation from the reference point~$\yref$. Here, $\Gp$ denotes the
matrix element of the pseudoscalar density between a pseudoscalar
state and the vacuum, and the arguments of $\Fp,\,\Gp$ refer to the
quark mass value at which the matrix elements have to be evaluated.
Using the definition of the current quark mass in terms of $\Fp,\,\Gp$
and $\mp$ (see, for instance, eqs.~(2.1) and~(2.2) in
ref.~\cite{mbar:pap3}), \eq{eq_RFG_def1} can be rewritten as
\bes
  \RFG(x) = \left(\frac{2y}{\mp^2(y)}\right)\left/
  \left(\frac{2\yref}{\mp^2(\yref)}\right)\right.
  = x\,\frac{\mp^2(\yref)}{\mp^2(y)}.
 \label{eq_RFG_def2}
\ees
Extracting the low-energy constants from the ratios $\RX,\,X={\rm
  M,\,F}$ instead of the masses and matrix elements themselves has
several advantages:
\begin{itemize}
\item The ratios $\RX$ can be computed with high statistical precision
  owing to the strong correlations between numerator and denominator;
\item The renormalization factors of the axial current and the
  pseudoscalar density drop out in $\RFG$ and $\RF$.\footnote{For
    $\Oa$ improved Wilson fermions, there remains a small
    mass-dependent renormalization proportional to
    $(\ba-\bp)(x-1)a\mref$ and $\ba(x-1)a\mref$ in the case of $\RFG$
    and $\RF$, respectively. Our experience has shown~\cite{mbar:pap3}
    that this can be safely neglected. In our calculations we set
    $\ba$ and $\bp$ equal to their one-loop perturbative
    values~\cite{impr:pap5}.}  The ratios can therefore be readily
  extrapolated to the continuum limit, so that all dependence on the
  lattice spacing is eliminated. Strictly speaking it is only in the
  continuum limit that one is justified to compare the predictions of
  ChPT with lattice data;
\item Since discretization errors are under good control in $\RX$ the
  effects of dynamical quarks can be isolated unambiguously.
\end{itemize}
The high level of statistical accuracy of the ratios is crucial in
order not to compromise the precision in the continuum
limit. Extracting the low-energy constants  from the $x$-dependence in
the continuum limit in turn guarantees that these estimates will not
be distorted by cutoff effects.

The simple renormalization and high precision of the ratios may also
be exploited to perform scaling tests for different fermionic
discretizations.

\subsection{Expressions for $\RFG$ and $\RF$ in ChPT \label{sec_RX}}

Below we give a list with the expressions for $\RFG$ and $\RF$ in
quenched and partially quenched ChPT. For simplicity we restrict
ourselves to the case of degenerate valence quarks. The case of
non-degenerate valence quarks in partially quenched ChPT is discussed
in detail in Appendix~\ref{sec_app1}.

Note that we have not yet specified the reference point~$\yref$. At
this stage the precise definition of its numerical value is not
needed, and we thus postpone its specification to
Section~\ref{sec_results}, where we describe our numerical results.

We begin by considering~$\RFG$ and~$\RF$ in quenched ChPT. By
inserting \eq{eq_MP_qchpt} into \eq{eq_RFG_def2} we obtain
\bes
  \RFG^{\rm q}(x) =
  \frac{1-(\delta-\frac{2}{3}\alpha_\Phi\yref)(1+\ln\yref) 
  +\yref[(2\alpha_8^{\rm q}-\alpha_5^{\rm q})-\frac{\alpha_\Phi}{3}]}
  {1-(\delta-\frac{2}{3}\alpha_\Phi x\yref)(1+\ln(x\yref))
  +x\yref[(2\alpha_8^{\rm q}-\alpha_5^{\rm q})
  -\frac{\alpha_\Phi}{3}]}. \label{eq_RFG_quen_rat}
\ees
Provided that all masses and couplings are small, it is allowed to
expand the denominator, which gives
\bes
  \RFG^{\rm q}(x) &=& 1+\delta\ln{x}
  -\textstyle\frac{2}{3}\alpha_\Phi\yref\left[
  x\ln{x}+(x-1)(\textstyle\frac{1}{2}+\ln\yref)\right] \nonumber\\
  & & -\yref(x-1)(2\alpha_8^{\rm q}-\alpha_5^{\rm q}).
  \label{eq_RFG_quen}
\ees
Similarly we obtain
\bes
  \RF^{\rm q}(x)=
  \frac{1+x\yref{1\over2}\alpha_5^{\rm q}}
       {1+\yref{1\over2}\alpha_5^{\rm q}}    \label{eq_RFPS_quen_rat},
\ees
which, after expanding the denominator, becomes
\bes
  \RF^{\rm q}(x)=
   1+\yref(x-1){1\over2}\alpha_5^{\rm q}.
  \label{eq_RFPS_quen}
\ees
In partially quenched ChPT it is useful to always define the reference
point $\yref$ at
\bes
  m_1=m_2=m_S=\mref,\quad \yref=\frac{2B_0\mref}{(4\pi F_0)^2}.
\label{eq_yref_PQ}
\ees
There are several possibilities to study the mass dependence of the
ratios $\RX$. Let us first consider the case of equal sea and valence
quark masses. The $x$-dependent part in $\RX$ is then obtained by
setting
\bes
  \hbox{SS:}\quad m_1=m_2=m_S=x\mref,
\label{eq_xdep_SS}
\ees
which will be labelled ``SS'' in the following. By taking the
appropriate limits in eqs.~(\ref{eq_yparam_PQ})
and~(\ref{eq_MP_pqchpt}) for the above choices of quark masses and
inserting the resulting expressions into the definition of $\RFG$, we
obtain, after expanding the denominator:
\bes
  \RFG^{\rm SS}(x) &=&
  1-\frac{1}{\Nf}\yref\left[x\ln{x}+(x-1)\ln{\yref}\right] \nonumber\\
  & &-\yref(x-1)\left[(2\alpha_8-\alpha_5)+\Nf(2\alpha_6-\alpha_4)\right].
  \label{eq_RFG_SS}
\ees
For $\RF$ the corresponding result is
\bes
   \RF^{\rm SS}(x)=
   1-\frac{\Nf}{2}\yref\left[x\ln{x}+(x-1)\ln\yref\right]
   +\yref(x-1)\textstyle\frac{1}{2}\left(\alpha_5+\Nf\alpha_4\right).
  \label{eq_RFPS_SS}
\ees
For $m_1,\,m_2\not=m_S$ the $x$-dependence can be mapped out using
either the valence or the sea quarks. In the former case, which we
label ``VV'' we define the $x$-dependent part through
\bes
   \hbox{VV:}\quad m_1=m_2=x\mref,\quad m_S=\mref
\label{eq_xdep_VV}
\ees
instead of \eq{eq_xdep_SS}, which leads to the expressions
\bes
\RFG^{\rm VV}(x) &=&
   1-\frac{1}{\Nf}\yref\left[(2x-1)\ln{x}+2(x-1)\ln\yref\right]
   \nonumber\\
   & &
   -\yref(x-1)\left[(2\alpha_8-\alpha_5)+\textstyle\frac{1}{\Nf}\right]
  \label{eq_RFG_VV}  \\
\RF^{\rm VV}(x) &=&
   1-\frac{\Nf}{4}\yref\left[
   (x+1)\ln\left(\textstyle\frac{1}{2}(x+1)\right)
   +(x-1)\ln\yref\right] \nonumber\\
   & &+\yref(x-1)\textstyle\frac{1}{2}\alpha_5.
   \label{eq_RFPS_VV}
\ees
A comparison of the expressions for $\RFG$ and $\RF$ for the ``SS''
and ``VV'' cases shows that they are sensitive to different linear
combinations of low-energy constants, depending on whether the
$x$-dependence is defined using~\eq{eq_xdep_SS} or~\eq{eq_xdep_VV}. In
particular, we find that it is possible to extract directly from
$\RFG^{\rm VV}(x)$ the linear combination $(2\alpha_8-\alpha_5)$,
which is relevant to the question of whether $\mup=0$.

There are several other possibilities to define the dependence of
$\RX$ on the quark masses, also allowing for non-degenerate valence
quarks. Details are listed in Appendix~\ref{sec_app1}.

\subsection{Extracting the low-energy constants \label{sec_extalpha}}

We now describe a method to determine the low-energy constants from
the ratios $\RX$ in the quenched and unquenched cases. To this end we
choose two values of mass parameters, $x_1,\,x_2$, and introduce the
quantity
\bes
  \Delta\RX(x_1,x_2) = \RX(x_1)-\RX(x_2),\quad x_1<x_2,\quad X={\rm
  M,\,F}. 
\label{eq_Delta_def}
\ees
By inserting the expressions for the ratios $\RX$ we can easily solve
for the low-energy constants. For instance, from
eqs.~(\ref{eq_RFPS_quen}) and~(\ref{eq_Delta_def}) we find
\bes
    \alpha_5^{\rm q} =
    2\frac{\Delta\RF^{\rm q}(x_1,x_2)}{(x_1-x_2)\yref}. 
\label{eq_alpha5_q}
\ees
Similar combinations of the $\alpha_i$'s can be obtained in partially
quenched QCD. The explicit expressions are given in
Appendix~\ref{sec_app1}, and one can easily convince oneself that they
serve to determine $\alpha_4,\,\alpha_5,\,\alpha_6$ and~$\alpha_8$.

The quantity $\Delta\RFG^{\rm q}$ is a special case, since it also
depends on the low-energy constants $\delta$ and $\alpha_\Phi$, which
only occur in the quenched theory. However, by inserting the estimates
for $\delta$ and $\alpha_\Phi$ quoted in the literature we can
solve for $(2\alpha_8^{\rm q}-\alpha_5^{\rm q})$, i.e.
\bes
  (2\alpha_8^{\rm q}-\alpha_5^{\rm q}) &=&
  \left\{\yref(x_1-x_2)\right\}^{-1} 
  \times\Big\{\delta\ln({x_1}/{x_2})-\Delta\RFG^{\rm q}(x_1,x_2)
  \nonumber \\
  & &-\textstyle\frac{2}{3}\alpha_\Phi\yref\left[
      x_1\ln{x_1}-x_2\ln{x_2} 
      +(x_1-x_2)\left(\textstyle\frac{1}{2}+\ln\yref\right)\right]
  \Big\}.
\label{eq_a8a5_q}
\ees
The expressions for $\RX$ themselves can also be used in order to
extract the low-energy constants from least-$\chi^2$ fits over a
suitably chosen interval in~$x$. The differences $\Delta\RX$, however,
have the advantage that some of the lattice artefacts may cancel.
Thus, instead of first extrapolating $\RX$ to the continuum limit and
then forming the differences $\Delta\RX$, one may compute
$\Delta\RX(x_1,x_2)$ at non-zero lattice spacing and then perform the
continuum extrapolation. Obviously the results must be independent of
the ordering of the two procedures, which offers an additional check
against the influence of lattice artefacts. By contrast, if a fitting
procedure is employed at non-zero lattice spacing then it is {\em a
  priori\/} not clear whether the continuum expressions for the
ratios~$\RX$ are valid.

Our method to extract the low-energy constants is only viable if there
is sufficient overlap between the region of validity of ChPT and the
range of quark masses accessible in current lattice simulations. On
present computers it is not possible to simulate very light quarks
without suffering from large finite-volume effects. Furthermore, the
fermion action and lattice spacings employed in this work do not allow
the use of quark masses which are below approximately half of the strange
quark mass. The reason is the occurrence of ``exceptional
configurations''~\cite{impr:pap3,excep:FNAL}, which correspond to
unphysical zero modes of the lattice Dirac operator. Therefore, since
one is restricted to a range of relatively large quark masses, one must
check the results against the influence of higher orders in the chiral
expansion: if large, these would modify the numerical estimates for
the low-energy constants considerably. This will be discussed in more
detail in Section~\ref{sec_results}.

\section{Continuum limit of $\RX$ and $\Delta\RX$ \label{sec_contlim}}

The ratios $\RX$ have been computed using the same quenched
configurations as in our earlier papers~\cite{mbar:pap2,mbar:pap3}.
Results for hadron and current quark masses, as well as for $\Fp$
and~$\Fp/\Gp$ are listed in Table~1 of ref.~\cite{mbar:pap3}.  Details
of our numerical procedures are described in Appendix~A of the same
paper. Here we only mention that non-perturbative $\Oa$ improvement
\cite{impr:sw,impr:pap1,impr:pap3} has been employed, and we will
therefore assume that the remaining discretization errors are of
order~$a^2$.

In contrast to our earlier papers the statistical errors in this work
have been estimated using a bootstrap procedure~\cite{efron}, with 200
bootstrap samples generated from the sets of gauge configurations.
This allows us to keep a constant number of bootstrap samples at every
value of the bare coupling, regardless of the number of
configurations. By performing the continuum extrapolation of $\RX(x)$
for every bootstrap sample, our error procedure thus preserves the
correlations in the mass parameter~$x$. Throughout this paper we quote
the symmetrized error from the central 68\% of the bootstrap
distribution.

\begin{figure}[tb]
\hspace{0cm}
\vspace{-1.cm}

\centerline{
\leavevmode
\psfig{file=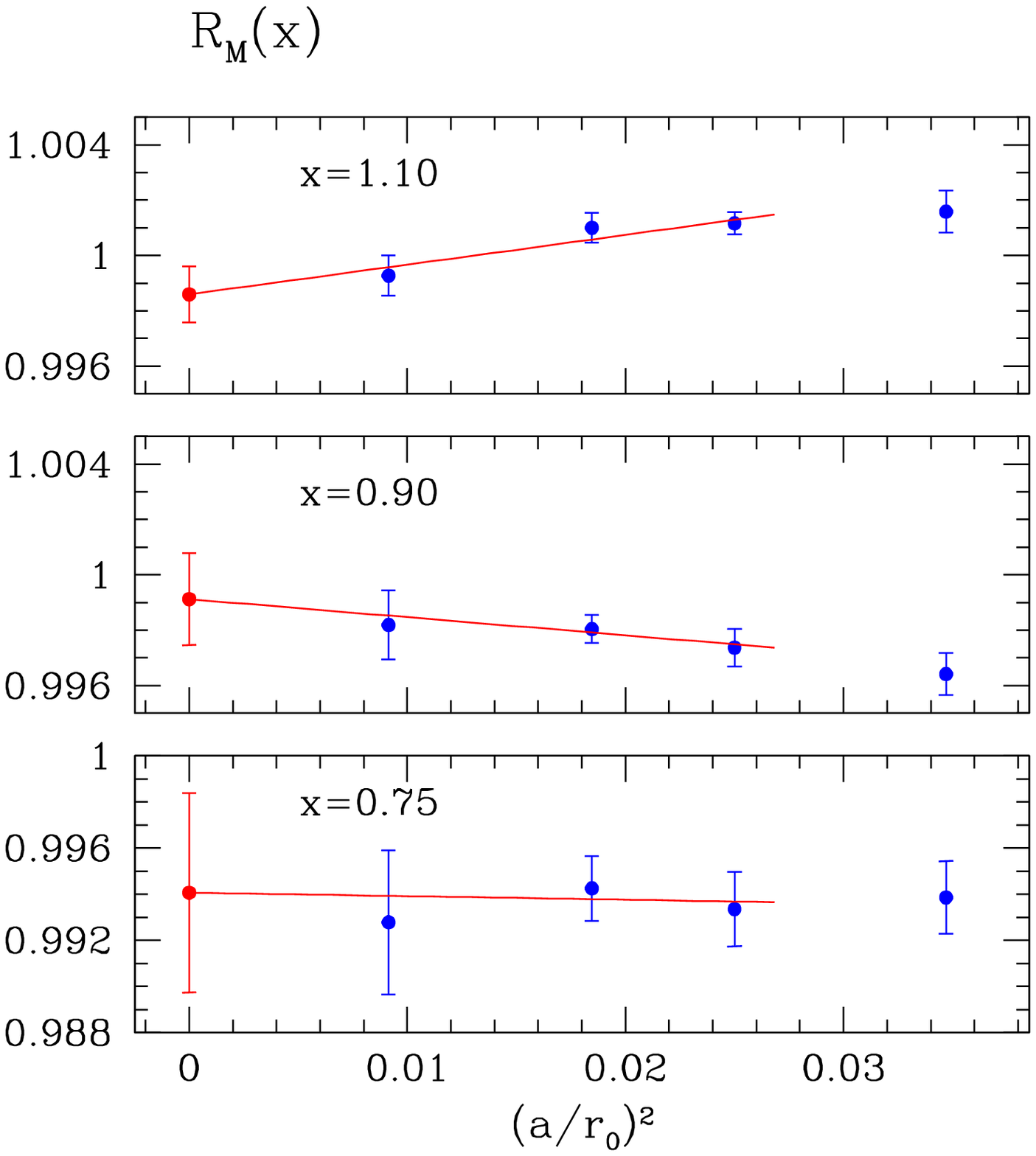,width=10cm}
\leavevmode
\hspace{-2.5cm}
\psfig{file=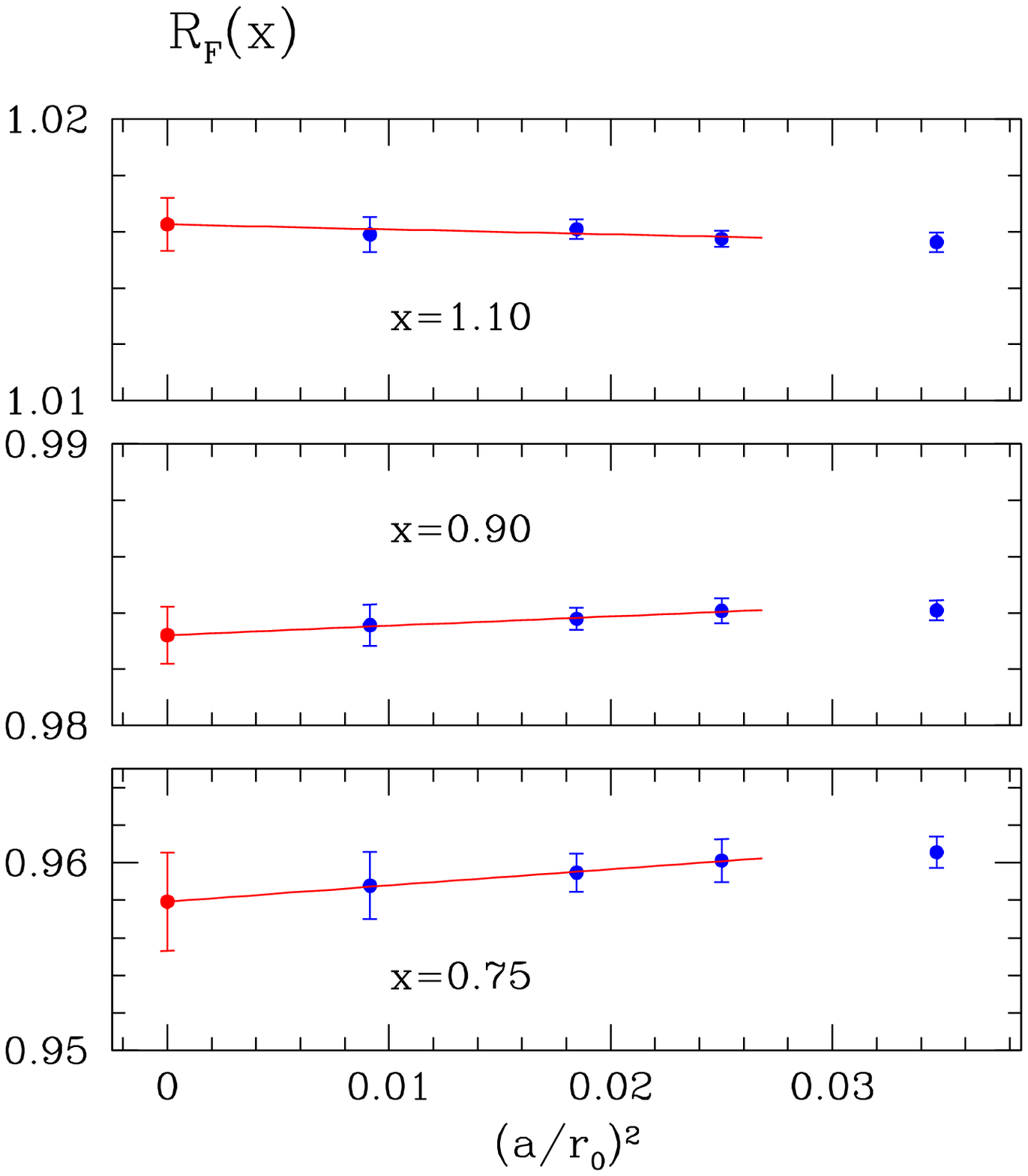,width=10cm}
}
\vspace{-0.5cm}
\caption{\footnotesize
  Continuum extrapolations of $\RFG(x)$ and $\RF(x)$ for selected
  values of the mass parameter~$x$.
\label{fig_cont_ext}}
\end{figure}

The value of the reference quark mass~$\mref$ is defined through
the condition
\bes
   \left(\mp r_0\right)^2\Big|_{m=\mref}=3,  \label{eq_mref_def}
\ees
a choice that has also been considered in
refs.~\cite{mbar:pap3,qspect:ukqcd99}. For $F_0=\Fpi=93.3\,\MeV$ the
numerical value of $\yref$ is thus $\yref=0.3398\ldots$. Lattice data
for the hadronic radius~$\r_0$~\cite{pot:r0} have been taken from
ref.~\cite{pot:r0_SU3}. For $r_0=0.5\,\fm$ the definition
of~\eq{eq_mref_def} corresponds to a pseudoscalar meson whose squared
mass is roughly twice as large as the kaon mass squared, and therefore
$\mref\approx\mstrange$ (with ``s'' standing for ``strange'').  The
results for $\Fp$ and $\Fp/\Gp$ obtained for several values of the
bare coupling have been interpolated in the current quark mass~$m$ to
common values of $x=m/\mref$. To this end the two neighbouring points
which straddle the value of~$x$ were used in a linear interpolation.
If~$x$ was slightly outside the range of simulated quark masses, a
linear extrapolation was performed using the three nearest points. The
stability of this procedure was checked by varying the input masses
and/or performing quadratic interpolations/extrapolations. Our sets of
simulated quark masses cover the range $0.75\leq{x}\leq1.4$, and we
have chosen increments of 0.05 to map out the mass dependence
of~$\RX(x)$.

\begin{figure}[bt]
\hspace{0cm}
\vspace{-5.cm}

\centerline{
\psfig{file=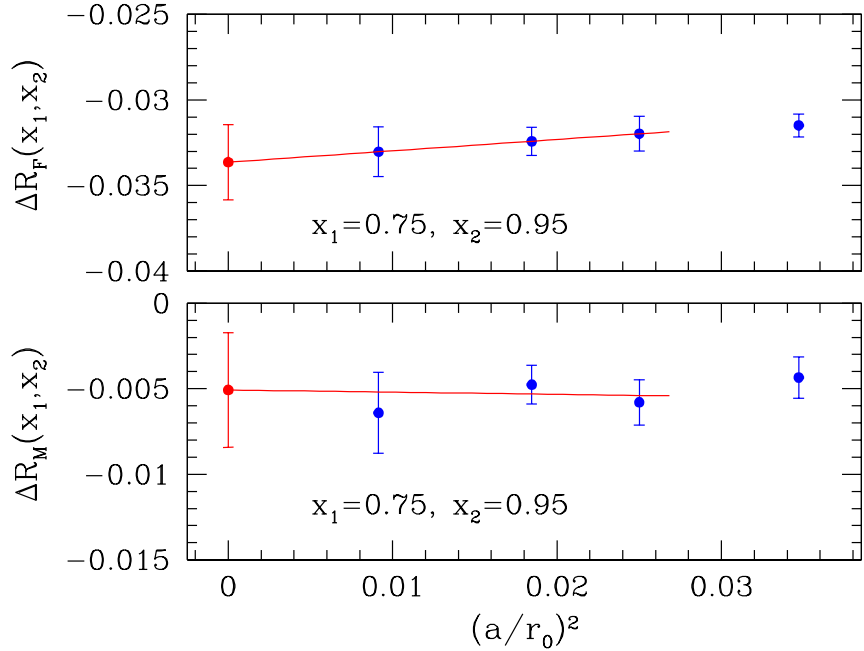,width=14cm}
}
\vspace{-0.5cm}
\caption{\footnotesize
  Continuum extrapolations of $\Delta\RFG$ and $\Delta\RF$.
\label{fig_del_cont_ext}}
\end{figure}

In \fig{fig_cont_ext} we plot the ratios $\RFG(x)$ and $\RF(x)$
against~$(a/r_0)^2$ for three different values of~$x$. The plots show
that lattice artefacts are very small in general and are consistent
with leading cutoff effects of order~$a^2$. As a safeguard against
higher-order lattice artefacts, we have excluded the points computed
for our coarsest lattice ($a\approx0.1\,\fm$) from the extrapolation.
Results obtained by performing the extrapolations for all four values
of the lattice spacing were entirely consistent, albeit with a smaller
statistical error.

At non-zero lattice spacing the statistical precision is better
than~0.3\% and~0.2\% for $\RFG$ and~$\RF$, respectively. In the
continuum limit these figures are only slightly larger, namely~0.4\%
for $\RFG$ and~0.3\% for~$\RF$. This demonstrates the high level of
precision that can be achieved in the continuum limit. Furthermore it
is clear that the extrapolation is well controlled.

As mentioned in Section~\ref{sec_stg} the results for $\Delta\RX$ in
the continuum limit can be obtained either directly from the continuum
values of~$\RX$ or from a continuum extrapolation of $\Delta\RX$
computed at non-zero lattice spacing. The latter extrapolations are
shown in~\fig{fig_del_cont_ext} for a particular choice of~$x_1$
and~$x_2$. As in the case of the ratios $\RX$ themselves the continuum
limit is easy to control.

\section{Results \label{sec_results}}

We can now compare our results for the ratios $\RX$ to the expressions
predicted by ChPT. Since the numerical data have been obtained in the
quenched approximation, we will focus on the determination of
$\alpha_5^{\rm q}$ and $(2\alpha_8^{\rm q}-\alpha_5^{\rm q})$.

\subsection{Low-energy constants in quenched ChPT}

The determination of $\alpha_5^{\rm q}$ from \eq{eq_alpha5_q} is
straightforward, since it only depends on the mass parameters $x_1$
and $x_2$. However, in order to compute $(2\alpha_8^{\rm
  q}-\alpha_5^{\rm q})$ from \eq{eq_a8a5_q} one must make a suitable
choice of $\delta$ and $\alpha_\Phi$. Here we are going to consider
two cases, namely
\bes
  \hbox{Q1:} & & \delta=0.12\pm0.02,\quad\alpha_\Phi=0
  \label{eq_Q1def} \\
  \hbox{Q2:} & & \delta=0.05\pm0.02,\quad\alpha_\Phi=0.5.
  \label{eq_Q2def}
\ees
The reasoning which led us to consider these two distinct sets of
parameters is described in Appendix~\ref{sec_app2}.

\begin{figure}[tb]
\hspace{0cm}
\vspace{-1.cm}

\centerline{
\leavevmode
\psfig{file=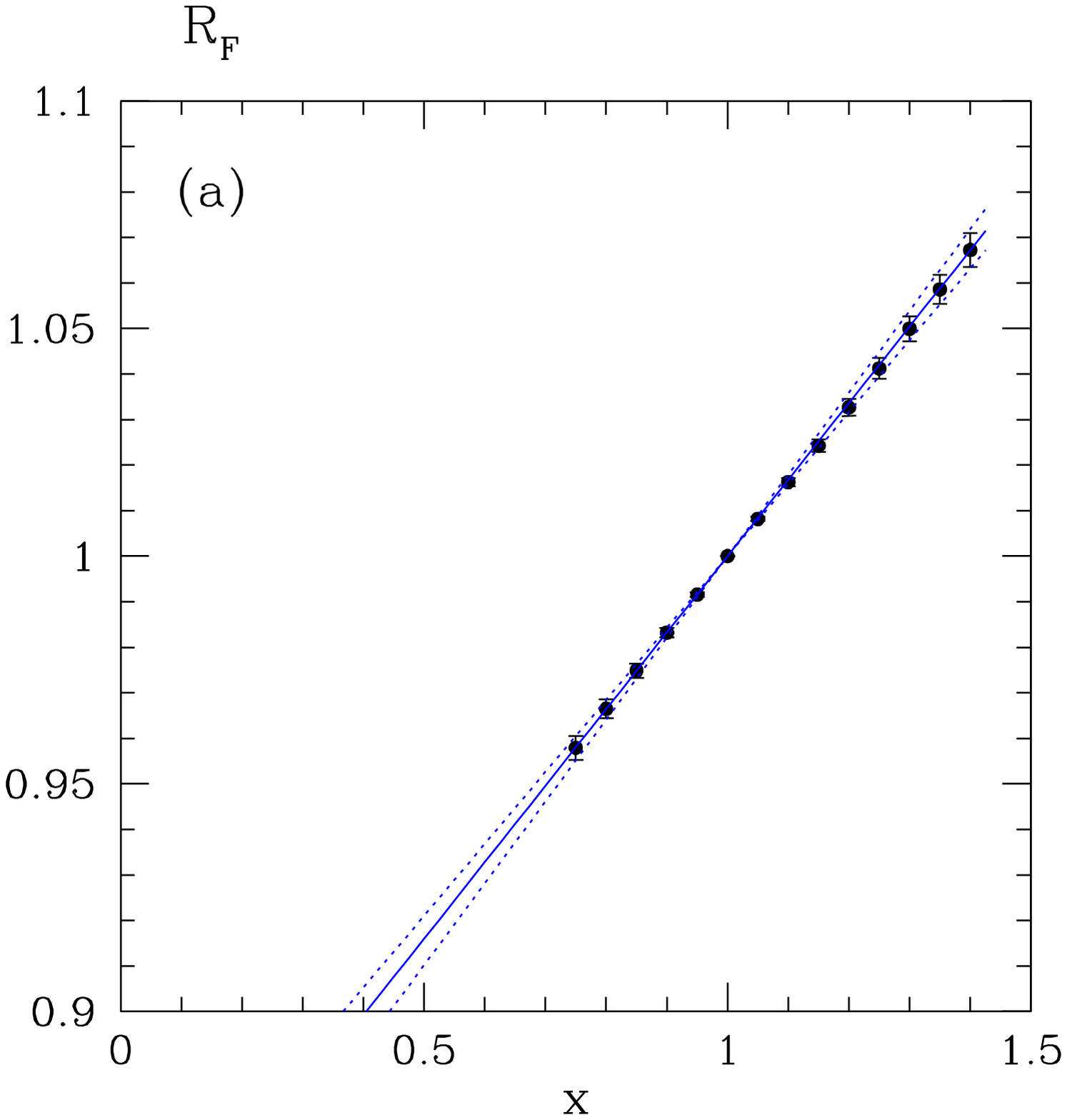,width=10cm}
\leavevmode
\hspace{-2.5cm}
\psfig{file=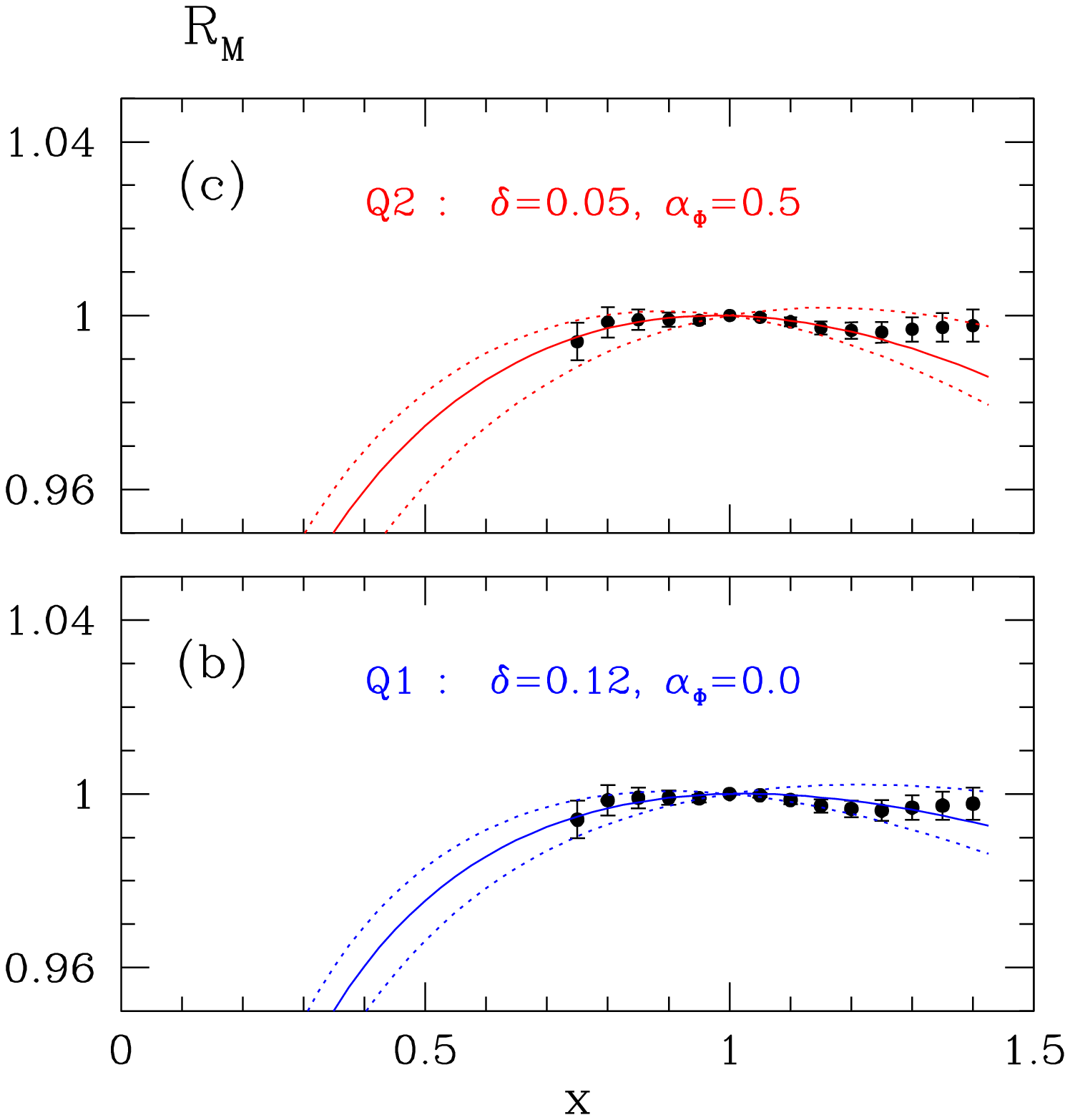,width=10cm}
}
\vspace{-0.5cm}
\caption{\footnotesize (a): Data for the ratio $\RF$ compared with the
  curve obtained by inserting the low-energy constant of
  eq.~(\protect\ref{eq_res_a5q}); (b) and (c): the same for the ratio
  $\RFG$ using the low-energy constants of
  eq.~(\protect\ref{eq_res_2a8a5q}) for parameter sets Q1 and Q2,
  respectively. Dotted lines indicate the variation due to the
  statistical uncertainty in the extraction of the low-energy
  constants.
\label{fig_RX_diff_err}}
\end{figure}

Our estimates for $\alpha_5^{\rm q}$ and $(2\alpha_8^{\rm
  q}-\alpha_5^{\rm q})$ have been obtained from $\Delta\RF(x_1,x_2)$
and $\Delta\RFG(x_1,x_2)$ for $x_1=0.75$ and $x_2=0.95$. This choice
was motivated by the desire to go to the smallest possible quark
masses, whilst maintaining a reasonably large $x$-interval in order to
check the stability of the results against variations in the
parameters~$x_1$ and~$x_2$.  Following this procedure, we obtained the
following estimates for the low-energy constants in quenched ChPT
\bes
  \alpha_5^{\rm q} &=& 0.99\pm0.06  \label{eq_res_a5q} \\
  (2\alpha_8^{\rm q}-\alpha_5^{\rm q}) &=& \left\{\begin{array}{lr}
         0.35\pm0.05\pm0.07; & \quad\hbox{Q1} \\
         0.02\pm0.05\pm0.07; & \quad\hbox{Q2}
                                      \end{array}.\right.
  \label{eq_res_2a8a5q}
\ees
Here, the first error is statistical, while the second (where quoted)
is due to the variation of $\pm0.02$ in the value of $\delta$ for both
parameter sets Q1 and Q2. These results can now be inserted into the
expressions for the ratios $\RFG$ and $\RF$, eqs.~(\ref{eq_RFG_quen})
and~(\ref{eq_RFPS_quen}), respectively, and the resulting curves are
compared with the data in Fig.~\ref{fig_RX_diff_err} (a)--(c).

The linearity of $\RF$ predicted by one-loop quenched ChPT is well
reproduced by the numerical data, resulting in very stable estimates
for $\alpha_5^{\rm q}$. Although the result in \eq{eq_res_a5q} has
been extracted from a fairly narrow $x$-interval, it provides a good
representation of the data over the entire range of quark masses
considered here. The determination of $(2\alpha_8^{\rm
  q}-\alpha_5^{\rm q})$ is also quite stable for both sets of values
of $\delta,\,\alpha_\Phi$, i.e. Q1 and Q2. In both cases a good
description of the numerical data is achieved for $x\;\lesssim\;1.2$,
with $(2\alpha_8^{\rm q}-\alpha_5^{\rm q})$ extracted for Q1 tracing
the data quite well even at the largest $x$-values.

Estimates for $\alpha_5^{\rm q}$ and $(2\alpha_8^{\rm q}-\alpha_5^{\rm
  q})$ obtained from the continuum extrapolations of $\Delta\RF$ and
$\Delta\RFG$ shown in Fig.~\ref{fig_del_cont_ext} are entirely
consistent with eqs.~(\ref{eq_res_a5q}) and~(\ref{eq_res_2a8a5q}). The
same is true if the low-energy constants are extracted directly from
fits to $\RX(x)$ for $x$ in the interval $0.75\leq{x}\leq0.95$; the
results are numerically almost identical to those obtained using the
differences $\Delta\RX$.

\subsection{Effects of higher orders in the quark mass}

Although the results presented in the last subsection suggest that the
one-loop formulae for quenched ChPT offer a good description of the
$x$-dependence of the ratios $\RX$, further work is needed to
corroborate these findings by extending the range of quark masses under
study to smaller values. For instance, the downward curvature observed
in the prediction for $\RFG$ when $x<0.75$ remains to be verified.

Furthermore, data at smaller quark masses will be required to
systematically assess the influence of higher-order terms in the
chiral expansion on the determination of the low-energy constants.
Such terms manifest themselves in additional contributions
proportional to $x^2$ in the formulae for $\RX$. Since the range
$0.75\leq{x}\leq0.95$ corresponds to pseudoscalar meson masses between
590 and $670\,\MeV$, it cannot be excluded that higher-order terms
have a sizeable impact on the extraction of the $\alpha_i$'s.

Even without results at smaller masses there are several possibilities
to study the effects of neglecting higher orders in ChPT on our
estimates for $\alpha_5^{\rm q}$ and $(2\alpha_8^{\rm q}-\alpha_5^{\rm
  q})$. We stress, however, that none of the methods described below
can replace the systematic investigation of smaller quark masses.

An obvious way to proceed is to add a quadratic term to the
expressions for $\RF^{\rm q}$ and $\RFG^{\rm q}$. For instance,
the modified expression for $\RF^{\rm q}$ reads
\bes
  \RF^{\rm q}(x) \longrightarrow
  1+\yref(x-1)\textstyle\frac{1}{2}\alpha_5^{\rm q}
   +\yref^2(x^2-1)\,d_{\rm F},
\ees
with a similar quadratic term proportional to $d_{\rm M}$ in the case
of $\RFG^{\rm q}$. One can now perform least-$\chi^2$ fits over the
{\it entire\/} range $0.75\leq{x}\leq1.4$, thereby extracting the
low-energy constants as well as $d_{\rm F},\,d_{\rm M}$. Because of
the linearity of the data for $\RF$ the only effect of including
$x^2$-terms in the determination of $\alpha_5^{\rm q}$ is to increase
its statistical error to $\pm\,0.19$. The estimates for
$(2\alpha_8^{\rm q}-\alpha_5^{\rm q})$ are more sensitive: compared
with eqs.~(\ref{eq_res_2a8a5q}) their central values increase by 0.11
and 0.22 for the parameter sets Q1 and Q2, respectively, while the
statistical error is increased to $\pm\,0.12$. The variation in the
central values, or the larger statistical error in the case of
$\alpha_5^{\rm q}$, may serve as an estimate of the uncertainty
induced by neglecting higher orders.

An alternative, albeit naive, estimate of the effect in question is
obtained by assuming that the chiral expansion converges like a
geometric series. This implies that one expects the quadratic
contributions to $\RX$ to be roughly as large as the square of the
linear ones. Here we consider $\RF^{\rm q}$ as the generic case,
since it does not involve logarithmic terms. Its linear contribution
amounts to $\approx16\%$ at $x=1$, so that the quadratic term is
estimated as $0.025\,x^2$. If we generalize this estimate, then the
systematic uncertainty in $\Delta\RX(x_1,x_2)$ due to neglecting
quadratic terms is given by
\bes
   \hbox{syst. error in}\;\Delta\RX(x_1,x_2) \approx
   \pm\,0.025\,(x_1^2-x_2^2). 
\label{eq_systerr_DRX}
\ees
Through eqs.~(\ref{eq_alpha5_q}) and~(\ref{eq_a8a5_q}) this is easily
translated into systematic errors\footnote{Instead of the constant
  0.025 in \protect\eq{eq_systerr_DRX} the reader may supply an
  alternative value, depending on whether our estimate is deemed too
  optimistic or pessimistic.} on $\alpha_5^{\rm q}$ and
$(2\alpha_8^{\rm q}-\alpha_5^{\rm q})$, as
\bes
   \hbox{syst. error in}\;\alpha_5^{\rm q} &\approx& \pm\,0.25 \\
   \hbox{syst. error in}\;(2\alpha_8^{\rm q}-\alpha_5^{\rm q})
   &\approx& \pm\,0.13,
\ees
which is of the same order of magnitude as the previous estimate.

Finally one can compare the expanded and unexpanded expressions for
$\RFG$ and $\RF$, which differ at order~$x^2$ (cf.
eqs.~(\ref{eq_RFG_quen_rat})--(\ref{eq_RFPS_quen})). By extracting the
low-energy constants from least-$\chi^2$ fits to the ratios in
eqs.~(\ref{eq_RFG_quen_rat}) and~(\ref{eq_RFPS_quen_rat}) we obtain
yet another set of results. For $\alpha_5^{\rm q}$ the central value
is larger by 20\%, whereas the result for $(2\alpha_8^{\rm
  q}-\alpha_5^{\rm q})$ remains essentially unchanged.

In order to present a synthesis of the various efforts to estimate the
uncertainty due to neglecting higher orders, we note that the typical
size of this systematic error amounts to $\pm\,0.2$ for both
$\alpha_5^{\rm q}$ and $(2\alpha_8^{\rm q}-\alpha_5^{\rm q})$. 

\subsection{Application: the ratio $\Fk/\Fpi$}

The result for $\alpha_5^{\rm q}$ extracted from $\RF^{\rm q}$ can
be used to compute the ratio of the decay constants of the kaon and
pion, $\Fk/\Fpi$. In fact, one usually employs the reverse procedure
by using the experimental result for $\Fk/\Fpi$ to extract the
phenomenological value of $\alpha_5$.

If we assume that contributions proportional to differences in the
valence quark masses can be neglected, we can simply use the
definition of $\RF^{\rm q}(x)$ to compute $\Fk/\Fpi$:
\bes
   \frac{\Fk}{\Fpi} = \frac{\RF^{\rm q}(\xK)}{\RF^{\rm q}(\xpi)}
   = 1+\yref(\xK-\xpi)\textstyle\frac{1}{2}\alpha_5^{\rm q}.
\label{eq_FKFpi_deg}
\ees
Here the dimensionless mass parameters $\xK$ and~$\xpi$ are related to
the kaon and pion masses by
\bes
   & & \xK \,\yref = \frac{\mk^2} {(4\pi F_0)^2} = 0.1782  \\
   & & \xpi\,\yref = \frac{\mpi^2}{(4\pi F_0)^2} = 0.0142,
\ees
where, as in ref.~\cite{mbar:pap3}, we have used $\mk=495\,\MeV$ and
$\mpi=139.6\,\MeV$. If the estimate for $\alpha_5^{\rm q}$ from
\eq{eq_res_a5q} is inserted we find
\bes
  & & \frac{\Fk}{\Fpi}=1.081\pm0.005\pm0.017.
\label{eq_FKFpi_deg_res}
\ees
Here the first error is statistical and the second is the estimated
uncertainty in $\alpha_5^{\rm q}$, due to neglecting quadratic terms. The
above result is significantly smaller than the experimental value of
${\Fk}/{\Fpi}=1.22$~\cite{PDG98}.

A formula for ${\Fk}/{\Fpi}$ in one-loop quenched ChPT, which also
accounts for differences in the valence quark masses, can be derived
from eqs.~(18) and~(20) of ref.~\cite{chir:GolLeuKpipi}. The
low-energy constant that appears in the one-loop counterterm is the
same as in the quenched degenerate
case~\cite{chir:ColPal97,ColPal_pcomm}. In our notation the full
one-loop expression for ${\Fk}/{\Fpi}$ reads
\bes
  \frac{\Fk}{\Fpi} &=& 
  1 +\,\yref(\xK-\xpi)\,\frac{1}{2}\,\alpha_5^{\rm q}
  -\frac{1}{2}\left\{\left(
  \delta-\frac{\alpha_\Phi}{3}\,\xK\,\yref\right) \right. \nonumber\\
  & & \left.
  -\,\frac{3\delta\,\xK-\alpha_\Phi\,\yref\xpi(2\xK-\xpi)}{6(\xK-\xpi)}
  \,\ln\left(\frac{2\xK-\xpi}{\xpi}\right)\right\}.
\ees
Note that one recovers \eq{eq_FKFpi_deg} when $\delta=\alpha_\Phi=0$.
For the two sets of parameters, Q1 and Q2, the results for
${\Fk}/{\Fpi}$ are evaluated as
\bes
  & & \frac{\Fk}{\Fpi} = \left\{\begin{array}{ll}
        1.125\pm0.005\pm0.016\pm0.007;\quad & \hbox{Q1} \\
        1.110\pm0.005\pm0.017\pm0.007;\quad & \hbox{Q2}
      \end{array},\right.
\label{eq_FKFpi_quen_res}
\ees
where the additional third error is due to the variation of
$\pm\,0.02$ in the input value for $\delta$. While contributions
proportional to differences in the quark masses enhance the result for
${\Fk}/{\Fpi}$ compared with \eq{eq_FKFpi_deg_res}, the value is still
smaller than experiment by about 10\,\%. Deviations of this order of
magnitude are typical of quantities computed in the quenched
approximation. This has previously been inferred from calculations of
the hadron spectrum~\cite{qspect:CPPACS,qspect:ukqcd99}, and the
results presented here firmly establish these findings for matrix
elements of local currents as well. The fact that ${\Fk}/{\Fpi}$ is
typically underestimated in quenched calculations has been observed
before~\cite{decay:GF11,ruedi_lat98}. Note, however, that our
estimates have much smaller uncertainties than those quoted in the
other references.

Finally we remark that the enhancement in the estimate of
${\Fk}/{\Fpi}$ due to differences in the quark masses, demonstrates
that these effects can be quite significant if the quark mass
difference is as large as that between the physical light and strange
quarks. By contrast, estimates of these effects based on masses in the
region of that of the strange quark tend to be much
smaller~\cite{mbar:pap3,qspect:ukqcd93}.

\subsection{Partially quenched ChPT}

In addition to our analysis of the quenched version of ChPT we can
also tentatively use the expressions for $\RFG$ and $\RF$ which are
derived in the partially quenched theory. This essentially serves two
purposes. On the one hand we can investigate to what extent the ratios
$R_X$ in quenched QCD are described by the formulae of partially
quenched ChPT. If a severe mismatch is encountered it will signal that
sea-quark effects are not accounted for. On the other hand, by
extracting the $\alpha_i$'s using the expressions of partially
quenched ChPT but quenched numerical data, we can test how much of the
relevant physical information for the low-energy constants in full QCD
is encoded in the mass dependence obtained in the quenched
approximation. The subsequent calculation of quantities such as the
ratio $\Fk/\Fpi$ and its comparison with experiment provides a
quantitative criterion for this task. In this spirit we have
investigated the cases labelled ``SS'' and ``VV'' discussed in
Section~\ref{sec_stg}.

If we employ the expression for $\RF^{\rm SS}$, \eq{eq_RFPS_SS}, we
find that the linearity of the numerical data is incompatible with the
additional logarithmic terms contained in the formula. As a
consequence an acceptable representation of the data for $\RF$ cannot
be achieved on the basis of \eq{eq_RFPS_SS}. We conclude that the mass
dependence of the quenched decay constant is {\em incompatible} with
the chiral behaviour predicted in full QCD!

By contrast, the expression for $\RFG^{\rm SS}$ fits the numerical
data very well over almost the entire range in~$x$. However, the
disagreement between the data and $\RF^{\rm SS}$ indicates that the
``SS'' formulae cannot be applied to the quenched data in any
reasonable manner. Therefore, we refrain from quoting an estimate for
$(2\alpha_8-\alpha_5)+3(2\alpha_6-\alpha_4)$, even though a good
modelling of the data for $\RFG$ can be achieved.

\begin{figure}[tb]
\hspace{0cm}
\vspace{-1.cm}

\centerline{
\leavevmode
\psfig{file=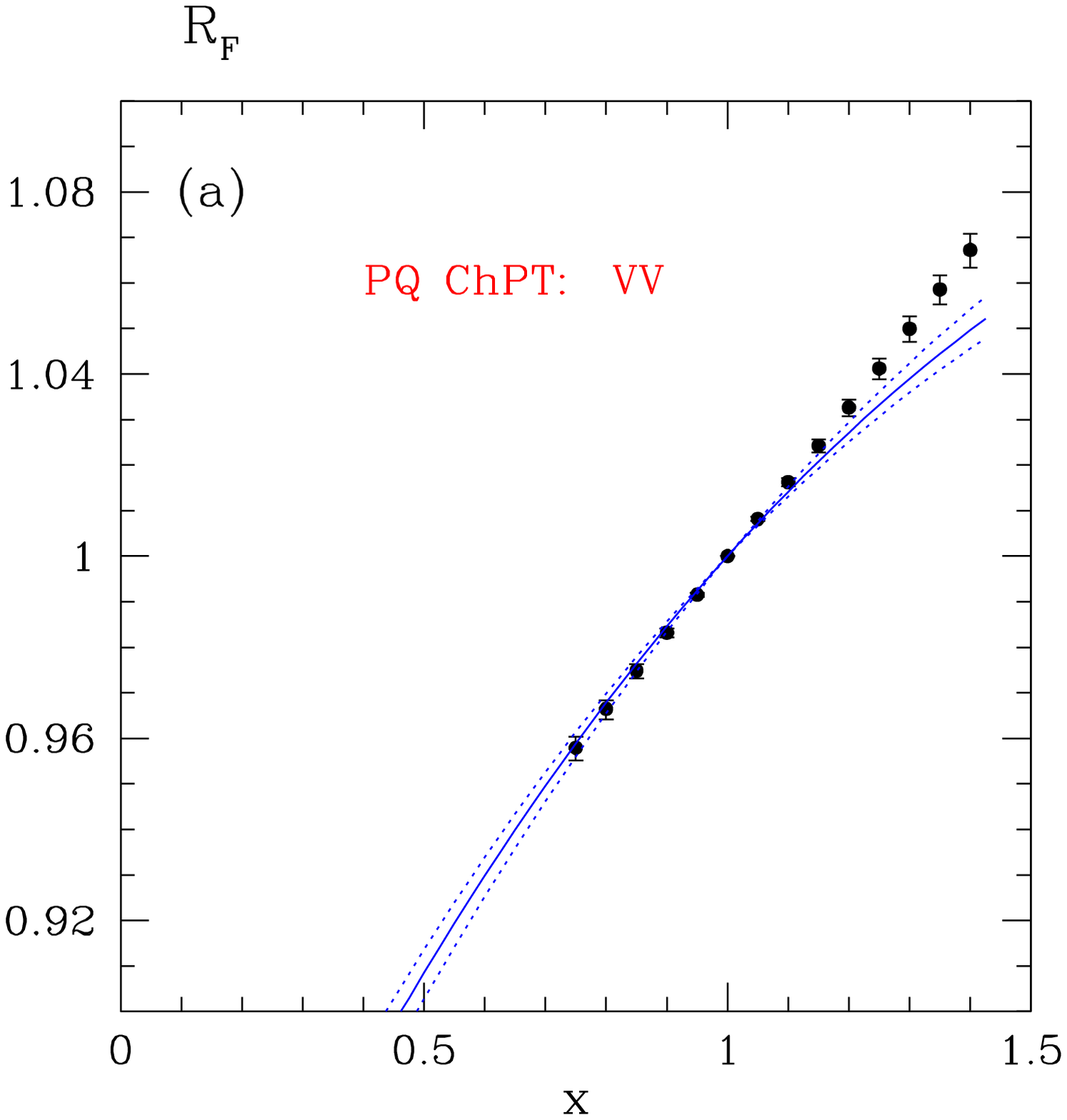,width=10cm}
\leavevmode
\hspace{-2.5cm}
\psfig{file=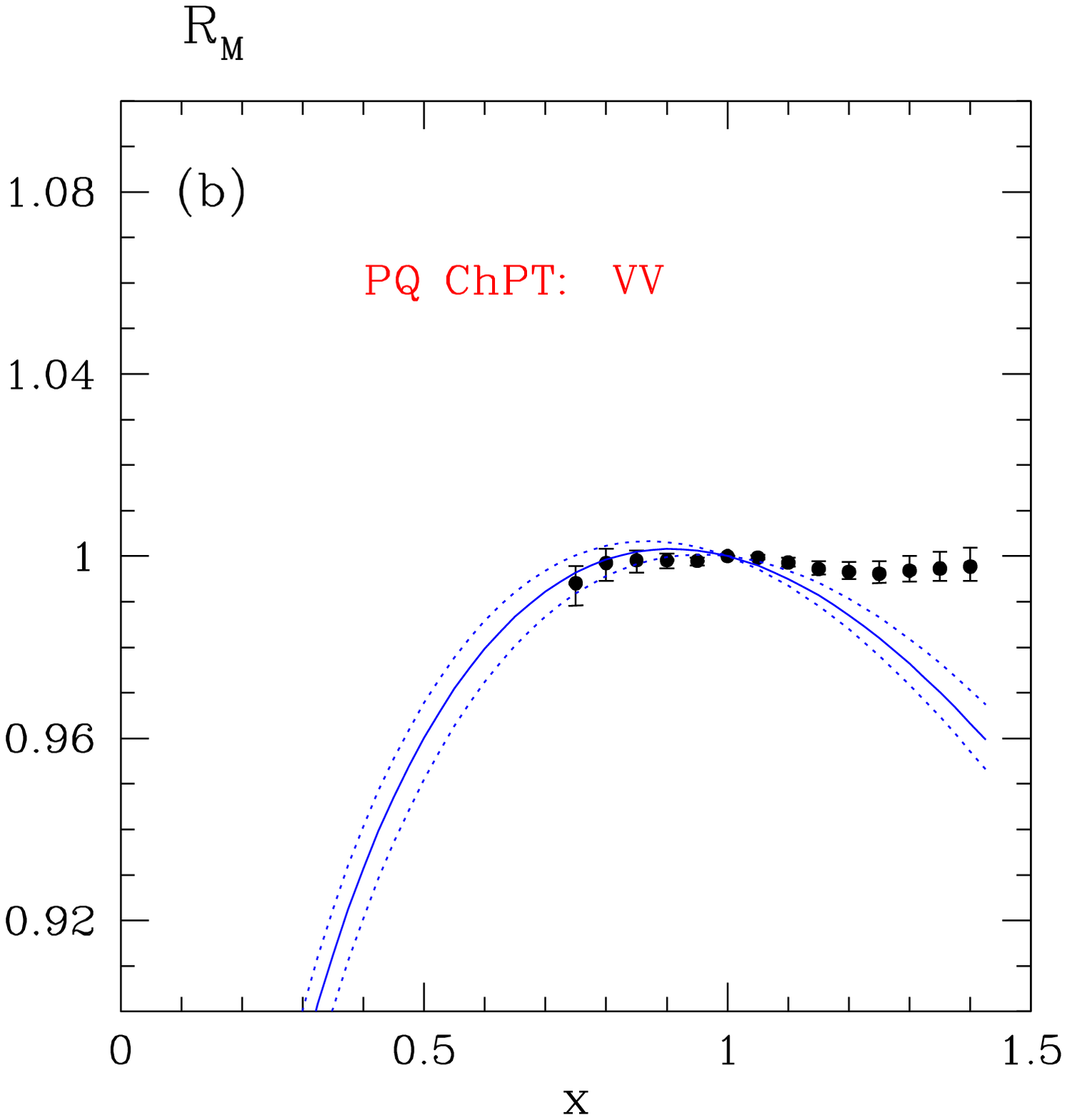,width=10cm}
}
\vspace{-0.5cm}
\caption{\footnotesize The ``VV'' case in partially quenched ChPT
  (a): data for the ratio $\RF$ compared with the curve obtained by
  inserting the low-energy constant of
  eq.~(\protect\ref{eq_res_a5_VV}); (b): the same for the ratio $\RFG$
  using the low-energy constants of
  eq.~(\protect\ref{eq_res_2a8a5_VV}). Dotted lines indicate the
  variation due to the statistical uncertainty in the extraction of
  the low-energy constants.
\label{fig_RX_diff_VV}}
\end{figure}

On the other hand, the expressions for both ratios obtained for the
``VV'' case do indeed lead to an acceptable representation of the
data, at least for $x\;\lesssim\;1$. Quoting only statistical errors,
the results for the low-energy constants read
\bes
  \alpha_5             &=& 0.75\pm0.06  \label{eq_res_a5_VV} \\
  (2\alpha_8-\alpha_5) &=& 0.15\pm0.05  \label{eq_res_2a8a5_VV}.
\ees
The curves corresponding to these values are shown in
Fig.~\ref{fig_RX_diff_VV} (a) and~(b). While for $x\;\lesssim\;1$ the
curves fit the data quite well, they deviate much more at larger~$x$
than the corresponding quenched ones. Compared with the quenched case,
the results for the $\alpha_i$ are roughly in the same ball park.

Using the value for $\alpha_5$ extracted from $\RF^{\rm VV}$ we can
now compute the ratio ${\Fk}/{\Fpi}$. Since the low-energy constants
in the partially quenched and full theories are the same, we have used
the expression derived in full QCD~\cite{chir:GaLe2}. In our notation
it reads
\bes
  \frac{\Fk}{\Fpi} &=& 1+\yref(\xK-\xpi)\textstyle\frac{1}{2}\alpha_5
  \nonumber\\ 
  & & +\,\frac{\yref}{4}\bigg\{
  (3\xpi-2\xK)\ln(\xpi\yref) -\xK\ln(\xK\yref) \nonumber\\
  & &
  \left. -\frac{1}{2}(4\xK-\xpi)\ln\left(\frac{4\xK-\xpi}{3\xpi}\right)
  \right\}.
\ees
After inserting the numerical values for $\xK, \xpi, \yref$ and
$\alpha_5$ one finds
\bes
   \frac{\Fk}{\Fpi} = 1.228\pm0.005\,\hbox{(stat)}\pm0.016.
\label{eq_FKFpi_VV}
\ees
Here the uncertainty of $\pm\,0.016$ is due to neglecting higher
orders. This result is actually compatible with the experimental
value, which is not entirely surprising, since our estimate of
$\alpha_5=0.75(6)$ agrees quite well with its phenomenological value
of $0.5\pm0.6$ quoted in \eq{eq_Li_stand}. These findings suggest that
the valence quark mass dependence of the numerically determined ratios
$\RX$ in the region of the strange quark mass is only weakly distorted
by neglecting dynamical quarks. The relevant quark mass effects which
account for the difference between the quenched result
\eq{eq_FKFpi_quen_res} and the experimental value $\Fk/\Fpi=1.22$,
appear to be due to the very light up and down quarks in full QCD.
Their non-linear effects can be efficiently accounted for by the
formulae of partially quenched ChPT.

\section{Summary and conclusions \label{sec_concl}}

In this paper we have proposed a general method to extract the
low-energy constants in effective chiral Lagrangians by studying the
mass dependence of suitably defined ratios of matrix elements and
matching it to ChPT. The ratios are typically obtained with high
statistical precision and can be extrapolated to the continuum limit
in a straightforward manner. Thus, the method offers a conceptually
clean way to determine the low-energy constants, since it is strictly
speaking only in the continuum limit that a comparison of lattice data
to ChPT can be performed.

The dimensionless ratios $\RX$ are also interesting in their own
right: they can be used to study the scaling behaviour of different
fermionic discretizations without having to compute renormalization
factors. Furthermore, the effects of dynamical quarks can be isolated
unambiguously, since lattice artefacts are under good control.

In this initial study we have presented results for the quenched
approximation. The typical value of the absolute statistical error of
$\pm\,0.05$ in the low-energy constants confirms that the achievable
precision is indeed quite high compared with ``conventional''
phenomenological determinations. The main limitation at present is
that results at smaller quark masses are not available.  It is
therefore not possible to systematically investigate whether there is
sufficient overlap between the range of simulated quark masses and the
domain of validity of ChPT. As a consequence the present estimated
uncertainty of $\pm\,0.2$ in the low-energy constants due to
higher-order terms are fairly coarse. It is expected, though, that the
simulation of smaller quark masses will be feasible within the
formulation of QCD with a ``twisted'' mass term
(tmQCD)~\cite{tmQCD:lat99}. In this construction the mass parameter
protects against unphysical zero modes of the Dirac operator, which
alleviates the problem of exceptional configurations encountered in
simulations with Wilson fermions. Initial studies have shown promising
results, and it is planned to exploit this method further in the
present context.

Returning to the problem of examining the question of whether
$\mup=0$, it is useful to recall that the problem can be solved
through a reliable determination of $\alpha_8$. In the quenched
approximation we obtain
\bes
  \alpha_8^{\rm q} = \left\{\begin{array}{lr}
         0.67\pm0.04\,\hbox{(stat)}\,\pm0.04; & \quad\hbox{Q1} \\
         0.50\pm0.04\,\hbox{(stat)}\,\pm0.04; & \quad\hbox{Q2}
                                      \end{array},\right.
\ees
where the second error is due to the variation in $\delta$, and the
additional uncertainty arising from neglecting higher orders is
estimated to be $\pm\,0.2$. There is {\it a priori\/} no reason why
the low-energy constant $\alpha_8^{\rm q}$ should be numerically close
to its counterpart in the full theory, and hence it would surely be
premature to give a full assessment of the problem on the basis of our
quenched results. Nevertheless, it is interesting to note that our
results for $\alpha_8^{\rm q}$ are quite close to the ``standard''
value for $\alpha_8$, \eq{eq_Li_stand}, which supports a non-vanishing
up-quark mass. By contrast, given our data for the ratios $\RX$ it
would be very difficult to accommodate a large negative value, which
is required for the up-quark to be massless (see \eq{eq_Li_m0}).

Finally we wish to add a few general comments on the philosophy of our
approach. Provided that the quark masses used in simulations are small
enough for the one-loop expressions of ChPT to apply, our procedures
altogether avoid chiral extrapolations of lattice results, which are
known to be quite difficult to control. The example of the computation
of $\Fk/\Fpi$ shows that the problem can be separated in two parts.
The information which involves quark masses near the chiral limit can
be extracted safely in ChPT, whereas the mass dependence for masses in
the region somewhat below the strange quark mass can be adequately
studied in lattice QCD. The latter part is the additional theoretical
input needed to determine some of the low-energy constants, which are
not accessible from chiral symmetry considerations alone. In short,
our method amounts to exploiting the complementary character of ChPT
and lattice QCD. The applicability of ChPT itself can be tested once
quark masses somewhat smaller than those in our present work become
accessible.

The next step is the application of our method to the case of
dynamical quarks. Here the relevant formulae that will be required are
listed in Appendix~\ref{sec_app1}. Simulations using either the O$(a)$
improved Wilson action for $\Nf=2$ flavours of dynamical
quarks~\cite{impr:csw_nf2} or other improvement
schemes~\cite{impr:Iwasaki} have already been
performed~\cite{dspect:ukqcd98,dspect:ukqcd_lat99,dspect:CPPACS99,dspect:CPPACS_lat99}.
Whereas an analysis of those results will be able to treat the
two-flavour case, the extension to the physically most interesting
case of $\Nf=3$ flavours will require a significant amount of
additional simulations. It is also worth investigating applications to
the case of non-degenerate sea quarks and to try to determine the
low-energy constant~$\alpha_7$~\cite{ShaSho_L7}. However, the main
message of this paper is the following. In order to settle the
question of whether $\mup=0$ one does not even require the same level
of accuracy as that of the quenched results presented here.  A
satisfactory analysis of the problem in the case of three degenerate
flavours is therefore quite a realistic prospect.

\vspace{1cm}
\noindent
{\bf Acknowledgements.} We are indebted to Gilberto Colangelo and
Elisabetta Pallante for essential clarifications concerning the
application of quenched ChPT. We are also grateful to Ruedi Burkhalter
for useful correspondence. This work is part of the ALPHA
Collaboration research programme. We thank DESY for allocating
computer time on the APE/Quadrics computers at DESY-Zeuthen and the
staff of the computer centre at Zeuthen for their support.

\vfill
\eject
\begin{appendix}

\section{Partially quenched ChPT -- the general case \label{sec_app1}}

In this appendix we list the expressions for $\RFG$ and $\RF$, as well
as for $\Delta\RFG$ and $\Delta\RF$, in partially quenched QCD. In
particular we discuss all possibilities to define the $x$-dependence
of $\RX$, allowing also for non-degenerate valence quarks.


We start by considering the expressions in
eqs.~(\ref{eq_yparam_PQ})--(\ref{eq_FP_pqchpt}). At the reference
point we require all quark masses to coincide with $\mref$
(cf.~\eq{eq_yref_PQ}). Two cases, labelled ``SS'' and ``VV'' have
already been discussed in Section~\ref{sec_RX}. In addition to ``VV'',
one can also study the $x$-dependence by varying the sea quark
mass~$m_S$ for fixed, degenerate valence quarks:
\bes
   \hbox{SS2:}\quad m_1=m_2=\mref,\quad m_S=x\mref.
\label{eq_xdep_SS2}
\ees
For non-degenerate valence quarks one may define
\bes
   \hbox{VS1:} & & m_1=x\mref,\quad m_2=m_S=\mref  \label{eq_xdep_VS1} \\
   \hbox{VS2:} & & m_1=\mref,\quad m_2=m_S=x\mref. \label{eq_xdep_VS2}
\ees
In order to list the expressions for $\RFG$ and $\RF$ for all cases
$\rm SS,\,VV,\ldots,VS2$ it is convenient to introduce the general
parameterization
\bes
  \RFG(x) &=& 1-\textstyle\frac{\yref}{\Nf}\rho_{\rm M}(x;\yref)
  -\yref(x-1)\lambda_{\rm M}(\mathbf{\alpha})   \\
  \RF(x)  &=& 1-\Nf\,\yref\,\rho_{\rm F}(x;\yref)-\yref(x-1)
  \lambda_{\rm F}(\mathbf{\alpha}).
\ees
Here, $\rho_{\rm M}$ and $\rho_{\rm F}$ are functions of $x$ and
$\yref$, and $\lambda_{\rm M}, \lambda_{\rm F}$ denote linear
combinations of the low-energy constants. The expressions for $\rho$
and $\lambda$ are shown in Tables~\ref{tab_rholam_RFG}
and~\ref{tab_rholam_RF}, respectively.

\begin{table}[tb]
\centering
\begin{tabular}{l l l}
\hline \\[-1.0ex]
  & $\rho_{\rm M}(x;\yref)$ & $\lambda_{\rm M}(\mathbf{\alpha})$
  \\[1.0ex]
\hline \\[-1.0ex]
SS  & $x\ln{x}+(x-1)\ln\yref$
    & $(2\alpha_8-\alpha_5)+\Nf(2\alpha_6-\alpha_4)$
    \\[1.0ex]
VV  & $(2x-1)\ln{x}+2(x-1)\ln\yref$
    & $(2\alpha_8-\alpha_5)+\frac{1}{\Nf}$
    \\[1.0ex]
SS2 & $-(x-1)\ln\yref$
    & $\Nf(2\alpha_6-\alpha_4)-\frac{1}{\Nf}$
    \\[1.0ex]
VS1 & $x\ln{x}+(x-1)\ln\yref$
    & $\frac{1}{2}(2\alpha_8-\alpha_5)$
    \\[1.0ex]
VS2 & 0
    & $\frac{1}{2}(2\alpha_8-\alpha_5)+\Nf(2\alpha_6-\alpha_4)$
    \\[1.0ex]
\hline
\end{tabular}
\caption{\footnotesize The functions $\rho$ and $\lambda$ for the
  ratio $\RFG(x)$.
\label{tab_rholam_RFG}
}
\end{table}

\begin{table}
\centering
\begin{tabular}{l l l}
\hline \\[-1.0ex]
  & $\rho_{\rm F}(x;\yref)$ & $\lambda_{\rm F}(\mathbf{\alpha})$
  \\[1.0ex]
\hline \\[-1.0ex]
SS  & $\frac{1}{2}\left(x\ln{x}+(x-1)\ln\yref\right)$
    & $-\frac{1}{2}(\alpha_5+\Nf\alpha_4)$
    \\[1.0ex]
VV  & $\frac{1}{2}\left(\frac{x+1}{2}\ln\frac{x+1}{2}
      +\frac{x-1}{2}\ln\yref\right)$
    & $-\frac{1}{2}\alpha_5$
    \\[1.0ex]
SS2 & $\frac{1}{2}\left(\frac{x+1}{2}\ln\frac{x+1}{2}
      +\frac{x-1}{2}\ln\yref\right)$
    & $-\frac{\Nf}{2}\alpha_4$
    \\[1.0ex]
VS1 & $\frac{1}{4}\left(\frac{x+1}{2}\ln\frac{x+1}{2}
      +\frac{x-1}{2}\ln\yref+\frac{1}{\Nf^2}\ln{x}\right)$
    & $-\frac{1}{4}(\alpha_5+\frac{1}{\Nf})$
    \\[1.0ex]
VS2 & $\frac{1}{4}\left(\frac{x+1}{2}\ln\frac{x+1}{2}
      +\frac{3(x-1)}{2}\ln\yref+\frac{\Nf^2-1}{\Nf^2}x\ln{x}\right)$
    & $-\frac{1}{4}\alpha_5-\frac{\Nf}{2}\alpha_4+\frac{1}{4\Nf}$
    \\[1.0ex]
\hline
\end{tabular}
\caption{\footnotesize The functions $\rho$ and $\lambda$ for the
  ratio $\RF(x)$.
\label{tab_rholam_RF}
}
\end{table}

Using the functions $\rho_X$ and $\lambda_X$, $X={\rm
  M,\,F}$, it is now quite easy to solve for particular linear
combinations of low-energy constants by considering the differences
$\Delta\RX(x_1,x_2)$ introduced in~\eq{eq_Delta_def}.

With these definitions the linear combinations of low-energy constants
denoted by $\lambda_X(\mathbf{\alpha}),\,X={\rm M,\,F}$ are simply
given by
\bes
  \lambda_{\rm M}(\mathbf{\alpha}) &=&
  -\frac{\Delta\RFG(x_1,x_2)+\frac{\yref}{\Nf}
  \left(\rho_{\rm M}(x_1,\yref)-\rho_{\rm M}(x_2,\yref)\right)}
  {\yref(x_1-x_2)}  \\
  \lambda_{\rm F}(\mathbf{\alpha}) &=&
  -\frac{\Delta\RF(x_1,x_2)+{\yref}\,{\Nf}
  \left(\rho_{\rm F}(x_1,\yref)-\rho_{\rm F}(x_2,\yref)\right)}
  {\yref(x_1-x_2)}.
\ees
By evaluating the right-hand side, using numerical data for
$\Delta\RX$, one can easily solve for the desired combination of
$\alpha_i$'s.

\section{Choices for $\delta$ and $\alpha_\Phi$ \label{sec_app2}}

\begin{figure}[tb]
\hspace{0cm}
\vspace{-5.cm}

\centerline{
\leavevmode
\psfig{file=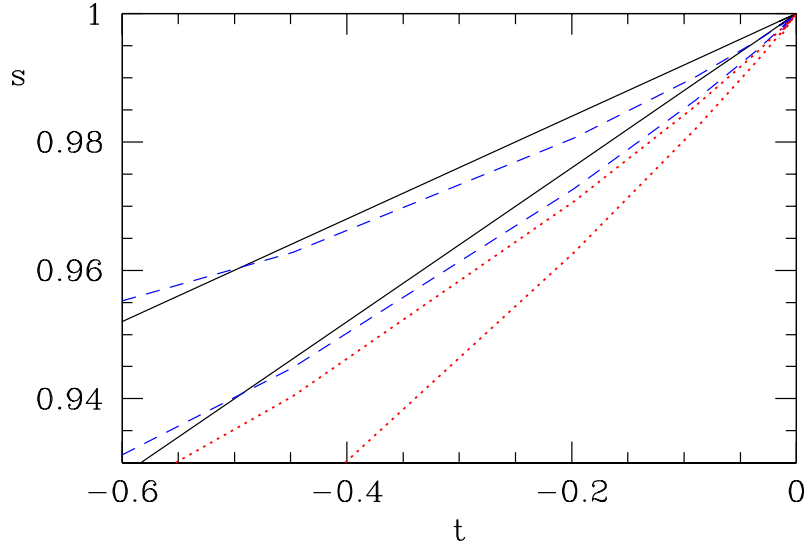,width=14cm}
\leavevmode
}
\vspace{-0.5cm}
\caption{\footnotesize The ``wedges'' defined through
  eq.~(\protect\ref{eq_wedge}) for different values of $\delta$ and
  $\alpha_\Phi$ as explained in the text.
\label{fig_wedges}}
\end{figure}

In this appendix we motivate our choices for the parameters $\delta$
and $\alpha_\Phi$ specified in eqs.~(\ref{eq_Q1def})
and~(\ref{eq_Q2def}). Here we will closely follow the analysis
presented by the CP-PACS Collaboration~\cite{qspect:CPPACS}, i.e. we
consider the ratio
\bes
  s=\frac{\mp^2(m_1,m_2)}{m_1+m_2}\cdot\frac{2m_1}{\mp^2(m_1,m_1)}
  \times\frac{\mp^2(m_1,m_2)}{m_1+m_2}\cdot\frac{2m_2}{\mp^2(m_2,m_2)},
\ees
where the arguments in $\mp^2$ have been included in order to
distinguish between degenerate and non-degenerate mesons. By inserting
the expressions for $\mp^2$ in quenched ChPT for degenerate and
non-degenerate quarks (cf. eq.~(9) in~\cite{chir:BerGol92}) one
obtains after expanding the denominator\footnote{Note that the
  one-loop counterterms drop out in the expanded version of $s$.}
\bes
  s = 1+\delta\left( 2-\frac{y_{11}+y_{22}}{y_{11}-y_{22}}
  \ln\frac{y_{11}}{y_{22}} \right) 
       +\frac{\alpha_\Phi}{3}\left(
  -(y_{11}+y_{22})+2y_{11}\frac{y_{22}}{y_{11}-y_{22}}\ln\frac{y_{11}}{y_{22}}
  \right). 
\label{eq_wedge}
\ees
In ref.~\cite{qspect:CPPACS} the quantity~$t$ was defined as
\bes
    t =
    2-\frac{y_{11}+y_{22}}{y_{11}-y_{22}}\ln\frac{y_{11}}{y_{22}}. 
\ees
For $\alpha_\Phi=0$ \eq{eq_wedge} reduces to a simple, linear relation
between $s$ and $t$:
\bes
   s = 1+\delta\cdot{t},
\ees
with the slope given by $\delta$. By plotting the numerically
determined values for the ratio~$s$ against~$t$ for $\alpha_\Phi=0$,
CP-PACS concluded that all their data were enclosed within the
``wedge'' defined by $\delta=0.08$--0.12. This wedge is represented in
Fig.~\ref{fig_wedges} by the solid lines.

So far the parameter~$\alpha_\Phi$ has not been considered in this
kind of analysis\footnote{Terms proportional to $\alpha_\Phi$ were
  also neglected in ref.~\cite{eta:FNAL_lat99}, where
  $\delta\approx0.08$ was quoted.}. We are now going to argue that a
value of $\alpha_\Phi\approx0.5$, as suggested by Sharpe
in~\cite{Sharpe_lat96}, is only compatible with the CP-PACS data
enclosed by the solid wedge, if $\delta$ is chosen in the range
$0.03$--0.07. To this end we identify $m_1=\mstrange$, so that
$y_{11}=0.343$. The mass ratio $y_{11}/y_{22}$ is then varied between
1 and 24.4, where the latter number is the value of
$2\mstrange/(\mup+\mdown)$ computed in standard
ChPT~\cite{leutwyler:1996}. For $\delta=0.08$--0.12, $\alpha_\Phi=0$
one thus recovers the wedge denoted by the solid line in
Fig.~\ref{fig_wedges}, which encloses the CP-PACS data.

If one combines $\delta=0.08$--0.12 with the independent estimate of
$\alpha_\Phi=0.5$ from~\cite{Sharpe_lat96} one obtains instead the
wedge defined by the dotted lines, which is clearly incompatible with
CP-PACS's numerical data. Agreement can only be restored if $\delta$
is lowered. The dashed curves, which correspond to
$\delta=0.03$--0.07, $\alpha_\Phi=0.5$, are again consistent with the
wedge defined by the solid lines. These observations lead us to
consider the two sets of parameters as specified in
eqs.~(\ref{eq_Q1def}) and~(\ref{eq_Q2def}).

\end{appendix}
\bibliography{biblist}        
\bibliographystyle{h-elsevier}   
\end{document}